\documentclass{optica-article}

\journal{opticajournal} % for journals or Optica Open

\articletype{Research Article}

\usepackage{lineno}
\usepackage[final]{changes}

\begin{document}

\title{Cavity-Enhanced Rydberg Atomic Superheterodyne Receiver}

\author{Yukang Liang,\authormark{1,$\dag$} Qinxia Wang,\authormark{1,3,$\dag$} Zhihui Wang,\authormark{1,2,$\dag$} Shijun Guan,\authormark{1} Pengfei Yang,\authormark{1,2} Yuchi Zhang,\authormark{1,4} Jun He,\authormark{1,2} Pengfei Zhang,\authormark{1,2} Gang Li,\authormark{1,2,*} and Tiancai Zhang\authormark{1,2}}

\address{\authormark{1}State Key Laboratory of Quantum Optics Technologies and Devices, Institute of Opto-Electronics, Shanxi University, Taiyuan 030006, China\\
\authormark{2}Collaborative Innovation Center of Extreme Optics, Shanxi University, Taiyuan 030006, China\\
\authormark{3}Institute for the History of Science and Technology, Shanxi University, Taiyuan 030006, China\\
\authormark{4}College of Physics and Electronic Engineering, Shanxi University, Taiyuan 030006, China\\
\authormark{$\dag$}The authors contributed equally to this work}

\email{\authormark{*}gangli@sxu.edu.cn} %% email address is required; see note below about the corresponding author designation

% use {asbstract*} to suppress the copyright line. Copyright information will be added in production

\begin{abstract*} 
High-sensitivity measurements of the microwave electric field are important in applications of communication and metrology. \replaced{The sensitivity of traditional Rydberg superheterodyne receivers in free space is effectively determined by the signal-to-noise ratio (SNR), which is often considered equivalent to sensitivity in practical sensing applications.}{The sensitivity of the traditional Rydberg superheterodyne receivers in free space is limited by signal-to-noise contrast.} In this work, we demonstrate a cavity-enhanced receiver, where an optical cavity significantly amplifies the interaction between the probe light and cesium atoms, which substantially improves the signal-to-noise ratio via enhancing the expansion coefficient \( \kappa \). \added{Here, $\kappa$ is the edge slope of the single peak obtained by fitting the double-peak EIT-AT spectrum, characterizing the response of the probe light to the frequency detuning of the coupling laser.}The sensitivity is thus boosted by a factor of approximately 19 dB. This study highlights the pivotal role of optical cavities in advancing Rydberg-based detection systems, offering a promising approach for high-sensitivity microwave electric field measurements.

\end{abstract*}

%%%%%%%%%%%%%%%%%%%%%%%%%%  body  %%%%%%%%%%%%%%%%%%%%%%%%%%
\section{Introduction}

The high-sensitivity detection of microwave ($\mu$-wave) electric fields is critical for various applications in communication \cite{jakes1994microwave}, radar technology \cite{assouly2023quantum}, and remote sensing \cite{lusch1999introduction}.  
One advantage of using Rydberg atoms over traditional antennas is their wide-range response from radio to light frequency, which is independent of sensor size, enabling a breakthrough beyond the Chu limit \cite{cox2018quantum, mao2022high, harrington1960effect}. 
In recent years, Rydberg atoms \cite{Gallagher_1994, adams2019rydberg, saffman2010quantum} have attracted intensive attention for $\mu$-wave detection due to their high sensitivity to external fields and intrinsic traceability \cite{holloway2017electric}, offering unprecedented sensitivity for electric field measurements. 
In 2007, electromagnetically induced transparency (EIT) was first observed in Rydberg atoms, paving the way for their use in quantum sensing \cite{mohapatra2007coherent}. 
The sensing of a 14-GHz $\mu$-wave field was demonstrated by utilizing EIT and Autler-Townes (AT) effects with an initial sensitivity of \(30 \, \mu \mathrm{V/cm/Hz^{1/2}}\), which was later improved to \(5 \, \mu \mathrm{V/cm/Hz^{1/2}}\) by using Mach-Zehnder interferometry \cite{sedlacek2012microwave,kumar2017atom}.
These early works established one classical principle for Rydberg atom-based $\mu$-wave sensing. 
Following the preceding works, various methods have been invented for high-sensitivity measurements of $\mu$-wave fields, including the use of Rydberg states with high principal quantum numbers \cite{Cai2023}, self-heterodyne frequency-comb techniques \cite{Dixon2023}, critical behavior in a many-body Rydberg atomic system \cite{ding2022enhanced}, stochastic switching in Rydberg systems \cite{he2020stochastic}, enhanced sensitivity in noisy environments \cite{Wu2024}, and the use of cold Rydberg atoms \cite{duverger2024metrology}.
Towards the real applications, Song et al. demonstrated digital $\mu$-wave communication via Rydberg-EIT detection with a rate of 500 kbps, verifying the feasibility of broadband PSK $\mu$-wave signal transmission \cite{Song2023_PhysRevApplied}. More intriguingly, Holloway et al. made the first attempt to apply quantum technology in the arts and demonstrated real-time guitar recording by using Rydberg atoms and EIT \cite{holloway2019real}.

The promotion of the sensitivity for Rydberg receivers is one of the most important tasks for the applications.
The sensitivity depends directly on the signal-to-noise ratio (SNR).
The SNR can be enhanced by using the atomic superheterodyne \cite{rohde1988communications, alabaster2012pulse}, in which the signal is selectively amplified.
Jing et al. first applied the superheterodyne in the Rydberg atomic receiver and improved the sensitivity to \(55 \, \mathrm{nV/cm/Hz^{1/2}}\) \cite{jing2020atomic}. 
Then, several improvements were made on other parameters, e.g., the bandwidth, the dynamic range, and the resolution, etc \cite{hu2023improvement, wu2023linear, wang2023noise, she2024rydberg}. 
The SNR can also be improved by directly enhancing the signal, which can be achieved by increasing the atom-light interaction. 
Optical cavity is one of the important devices to enhance the interaction between light and atoms by facilitating multiple reflections \cite{kimble1998strong, dutra2005cavity}. 
Peng and Wang et al. have demonstrated that enhancing the spectral contrast of Rydberg atoms can improve the sensitivity of $\mu$-wave electric field measurements through cavity-enhanced technology \cite{peng2018cavity, li2022enhanced, 8-20230039}. 
In our current study, we explore the application of cavity-enhanced techniques in Rydberg atomic superheterodyne receivers.
We show that the measurement sensitivity was improved by approximately \(19 \, \mathrm{dB}\) compared to the free-space mode by adopting an optical cavity.
The highest sensitivity of \( 176 \, \mathrm{nV/cm/Hz^{1/2}} \) is demonstrated in the current setup, and it can be improved further with proper cavity configuration.

\section{Experimental Setup}

\begin{figure}[htbp]
\centering\includegraphics[width=13cm]{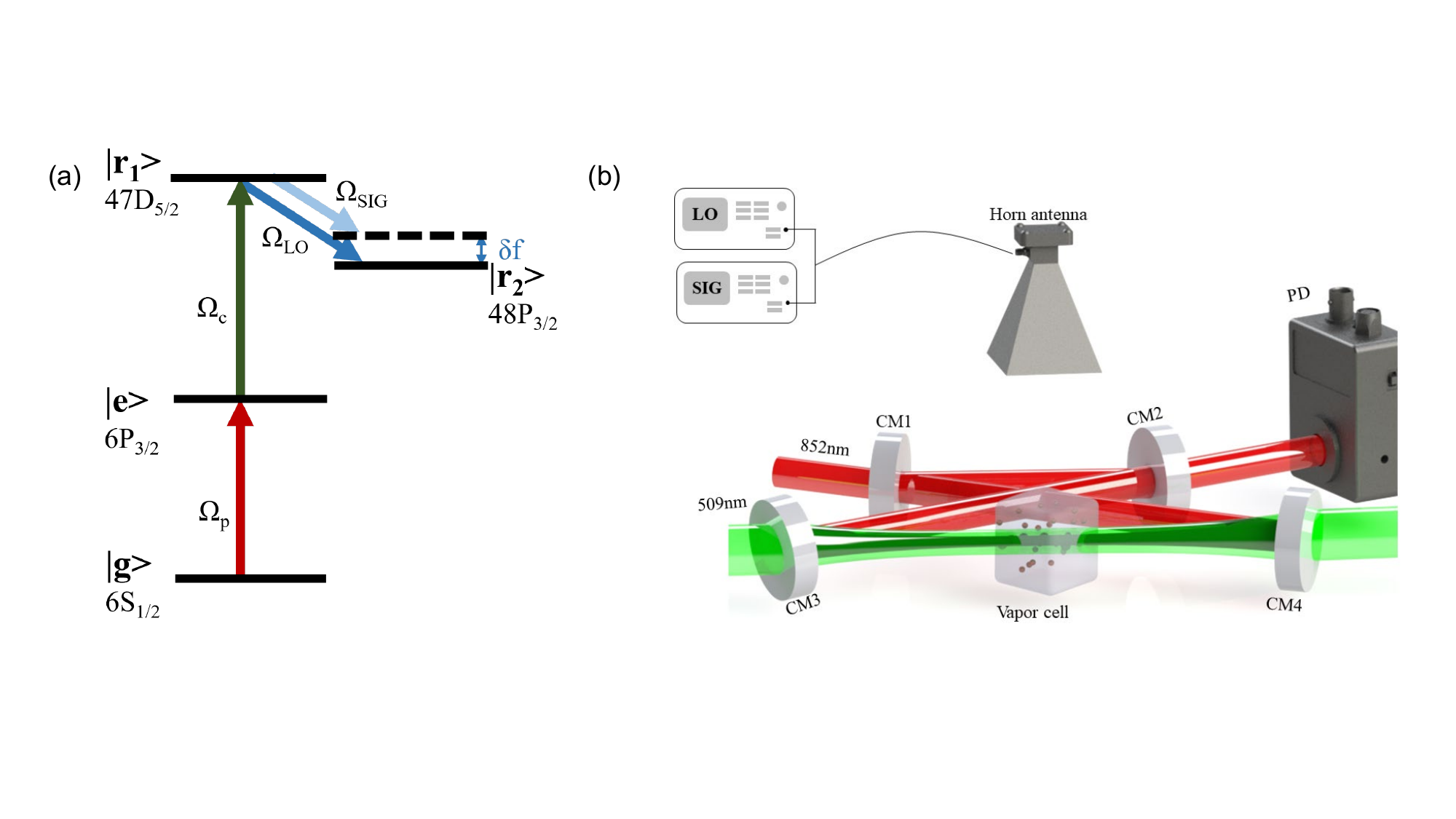}
\caption{(a) Energy levels of the $^{133}$Cs Rydberg receiver. The probe laser excites the atoms from the ground state \( \left| g \right\rangle \) to the intermediate state \( \left| e \right\rangle \), while the coupling laser excites them further to the Rydberg state \( \left| r_1 \right\rangle \). The LO and signal $\mu$-wave field couple the Rydberg states \( \left| r_1 \right\rangle \) and \( \left| r_2 \right\rangle \). (b) Experimental setup of the cavity-enhanced system, showing the optical cavity, vapor cell, and $\mu$-wave horn antenna.}
\label{fig:setup}
\end{figure}

The experimental setup is shown in Fig. \ref{fig:setup}, where an optical cavity with bow-tie configuration is employed to enhance the interaction between the probe light and cesium atoms. The cavity consists of four mirrors with a total length of 480 mm. 
The input (CM1) and output (CM2) mirrors are flat, having reflectivities of \(95\%\) and \(98\%\) at 852 nm. The other two mirrors (CM3, CM4) are plano–concave with a radius of curvature of 100 mm, designed for high reflectivity (\(>99.9\%\)) at 852 nm and high transmissivity (\(\sim90\%\)) at 509 nm, respectively. 
The finesse of the empty cavity is 85, and it \replaced{degrades}{degenerates} to 20 after the atom vapor cell is placed inside.
The beam waists of the probe and coupling lasers are located between CM3 and CM4 with sizes of \(50 \, \mu\mathrm{m}\) and \(80 \, \mu\mathrm{m}\), respectively. 
The cavity is actively stabilized to the frequency of the probe laser (852 nm) by using a frequency-stabilized 840-nm laser, allowing precise control of the cavity frequency to optimize the enhancement. A cesium vapor cell, with a length of 1 cm, is placed between CM3 and CM4 and the center is in coincidence with the waist of the cavity mode.

In the experimental setup, a near-infrared probe beam (852 nm) and a green coupling beam (509 nm) counter-propagate through the cesium vapor cell. The probe laser frequency is locked to the \(|6S_{1/2}\rangle \rightarrow |6P_{3/2}\rangle\) transition of cesium, while the coupling laser is tuned to the frequency of \(|6P_{3/2}\rangle \rightarrow |47D_{5/2}\rangle\) transition. 
A strong $\mu$-wave field with a frequency of 6.947 GHz works as a local oscillator (LO) which couples the Rydberg states \(|47D_{5/2}\rangle\) and \(|48P_{3/2}\rangle\), generating an AT splitting in the EIT transmission spectrum. 
Additionally, a weak test $\mu$-wave signal, detuned from the LO by \(\delta f = 150\) kHz, is combined with the LO and incident on the cesium vapor cell through a $\mu$-wave horn on the top. 
If we assume that the electric amplitudes for the LO and signal $\mu$-wave fields are \( E_{\text{LO}} \) and \( E_{\text{SIG}} \), respectively. 
At the condition \( E_{\text{LO}} \gg E_{\text{SIG}} \), the detected signal on the photodiode will be \cite{jing2020atomic}
\begin{equation}
    P_{\text{out}}(t) = \left| P(\delta f) \right| \cos(2\pi \delta f t + \delta \varphi)
\end{equation}
with \( \delta f\) and \( \delta \varphi \) the frequency and phase difference between the LO and signal $\mu$-wave fields. 
The amplitude \( \left| P(\delta f) \right| \) has a relation with the Rabi frequency of the test signal field \( \Omega_{SIG} \) as
\begin{equation}
    \left| P(\delta f) \right| = \left|\Omega_{SIG}\kappa\right|=\left| \frac{\sqrt{2} \mu_{MW}E_{SIG}\kappa }{\hbar } \right|.
\end{equation}
\( \kappa \) is an expansion coefficient, which is the edge slope of the single peak in the double-peak EIT-AT spectrum. \( \mu_{\text{MW}} \) is the electric dipole moment between the two Rydberg states.
The transmitted probe light is detected by a photodiode, which converts the intensity modulation into an electrical signal. 
The signal is subsequently analyzed using a spectrum analyzer. 
Our experimental setup enables precise comparison between free-space and cavity-enhanced Rydberg receivers: the setup can be converted to the free-space Rydberg receiver just by blocking the optical path between CM1 and CM2.

\section{Experimental Results}

\begin{figure}[htbp]
\centering\includegraphics[width=13cm]{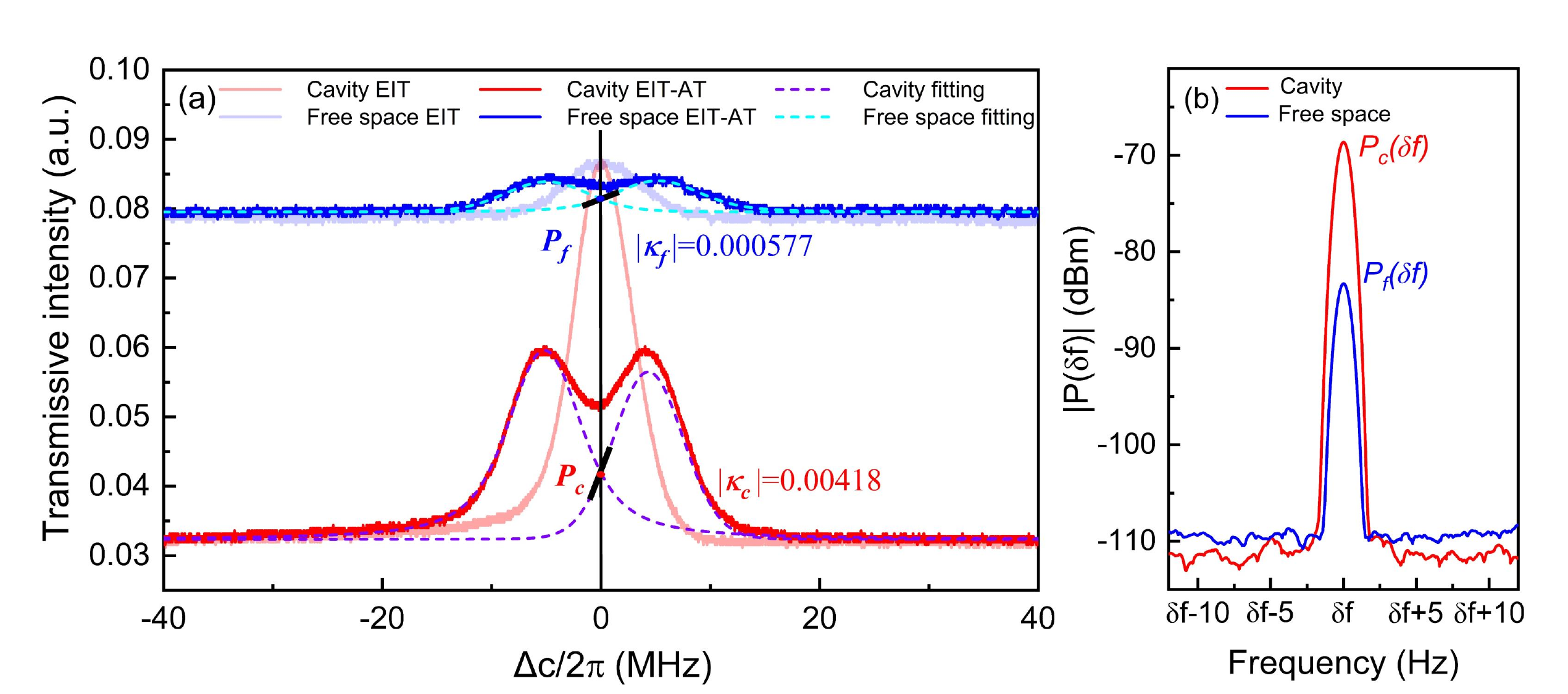}
\caption{(a) Comparison of EIT-AT splitting spectra and expansion coefficient \( |\kappa| \) in cavity-enhanced and free-space cases. Cyan and purple curves represent the Voigt double-peak fittings for the two cases. The single-peak edge slopes at the resonance are defined as \( |\kappa_c| \) for the cavity-enhanced case and \( |\kappa_f| \) for the free-space case, respectively. The experimental parameters are: $P_{852} = 1.22$ $\mu$W (free-space), $P_{852} = 0.85$ $\mu$W (cavity), $P_{509} = 6.84 $ mW, and \( E_{\text{LO}} = 2.04 \, \mathrm{mV/cm}\).
(b) The signals acquired by the spectrum analyzer in the cavity-enhanced (red) and the free-space (blue) cases with $E_{\text{SIG}} = 31.3$ $\mu$V/cm, where the resolution bandwidth (RBW) is set to 1 Hz.}
\label{fig:eit_at_spectra}
\end{figure}

In our experiment, the EIT spectra were observed by scanning the frequency of the coupling laser.
The strong LO field induces a splitting of the Rydberg energy levels, forming the AT splitting spectrum. By locking the coupling laser frequency to the resonance point of the \( \left| e \right\rangle\ \rightarrow \left| r_1 \right\rangle\) transition, the Rydberg atoms acted as an optimized mixer under the influence of the local oscillator field.
\replaced{Figure~\ref{fig:eit_at_spectra}(a) shows the typical EIT-AT splitting spectra under cavity-enhanced and free-space cases, with transmissive intensity of the probe light varying as a function of the frequency detuning (\( \Delta_c \)) of the coupling light under the same conditions.}{Figure~\ref{fig:eit_at_spectra}(a) shows the typical EIT-AT splitting spectra under cavity-enhanced and free-space cases under the same conditions.} The slopes \(|\kappa_c|\) and \(|\kappa_f|\) were extracted from single-peak fits at the resonance points \added{(\(P_{c}\) and \(P_{f}\))} in the cavity-enhanced and free-space cases, respectively. The experimental results demonstrate that the expansion coefficient in the cavity-enhanced case, \( |\kappa_c| = 0.00418 \) \added{(2\(\pi\) *a.u./MHz)}, is significantly larger than that in the free-space case, \( |\kappa_f| = 0.000577 \) \added{(2\(\pi\) *a.u./MHz)}, indicating an improvement of approximately 18 dB. 
This confirms that the optical cavity can enhance the interaction between the probe light and the atoms.
The boost of the expansion coefficient will finally improve the detection sensitivity of the system dramatically.
Figure~\ref{fig:eit_at_spectra}(b) compares the detected signal under cavity-enhanced and free-space cases, as measured directly by the spectrum analyzer. The peak intensity in the cavity-enhanced case is obviously higher than that in the free-space case, with an enhancement of approximately 18 dB. The noise base is also 2 dB lower than the free-space case. Therefore, a much higher SNR can be obtained by using a cavity.

\begin{figure}[htbp]
\centering\includegraphics[width=13cm]{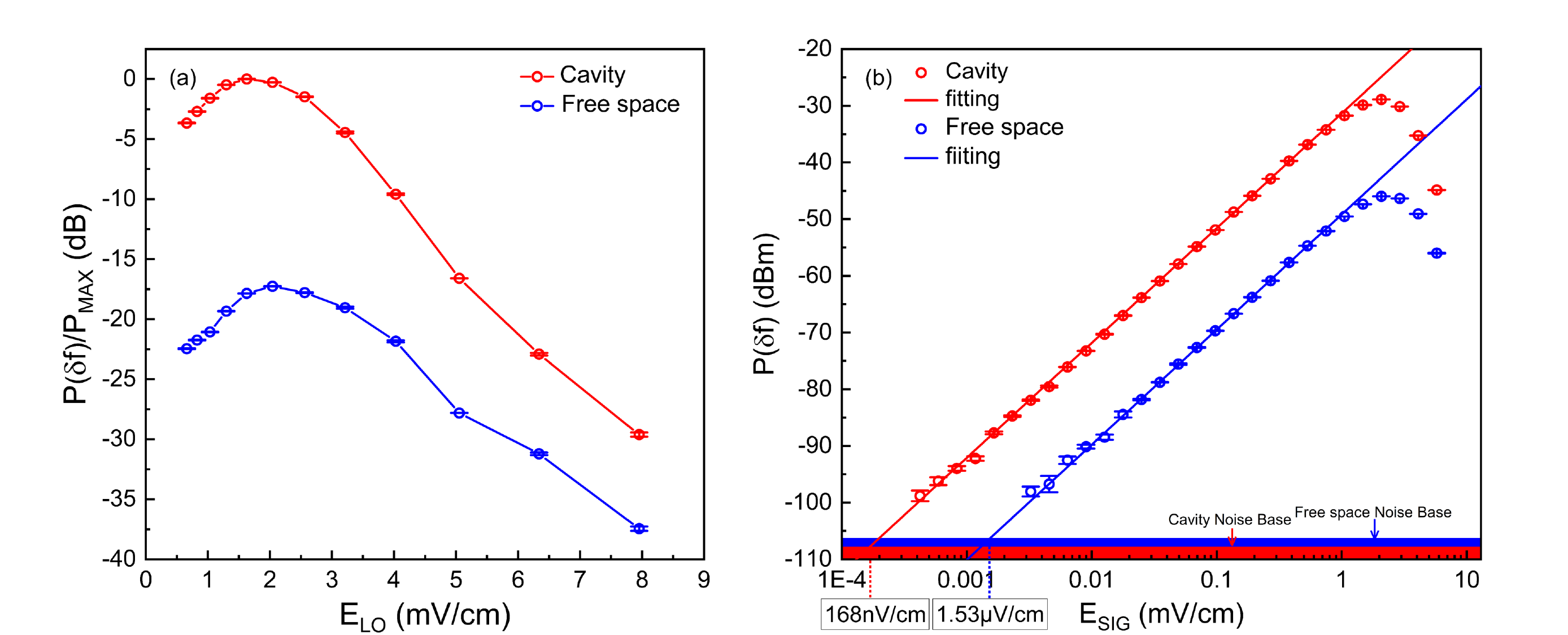}
\caption{(a) The measured electric signal \( P(\delta f)/P_{\text{MAX}} \) of the atomic superheterodyne system as a function of the LO field intensity \( E_{\text{LO}} \). The data in the cavity-enhanced case are represented by red circles, while the data in the free-space case are represented by blue circles. 
(b) The relationship between the measured electric signal of the atomic superheterodyne system and the electric field of the signal $\mu$-wave under cavity-enhanced case and free-space case. Red and blue circles represent the average values for independent 100 measurements for the cavity-enhanced case and free-space case, respectively. Error bars indicate the standard deviation of the measurements. Red and blue lines represent the linear fits in their linear regions. The red and blue shaded areas at the bottom represent the noise base directly read by the spectrum analyzer for the cavity-enhanced case and free-space case, respectively.}
\label{fig:discussion}
\end{figure}

The amplitude of the intermediate-frequency (IF) signal generated by atomic mixing depends on the intensity of the LO field. 
By adjusting the amplitude of the LO field \( E_{\text{LO}} \), the system can be optimized to its optimal working point, where the output signal amplitude reaches its maximum. Figure~\ref{fig:discussion}(a) shows the variation of the normalized signal amplitude \( P(\delta f)/P_{\text{MAX}} \) with the value of \( E_{\text{LO}} \) in both cavity-enhanced and free-space cases. 
We can see that the signals in the cavity-enhanced case are much higher than those in the free-space case.
The data show that the two cases exhibit different optimal working points. The optimal working points are 2.04 and 1.65 $\mathrm{mV/cm}$ for the free-space and cavity-enhanced cases, respectively.
The maximum signal of the cavity-enhanced case is 19 dB higher than that in the free-space case.
These results validate the advantages of cavity enhancement in reducing the minimum detectable electric field intensity and improving sensitivity. 

To systematically evaluate the performance of the cavity-enhanced Rydberg receiver, the dependences of the measured signal on the applied $\mu$-wave are investigated for both the cavity-enhanced and free-space cases. The results are displayed in Fig~\ref{fig:discussion}(b), where the powers of the probe and coupling lights are set as \(1.35 \, \mu\text{W} \) and \( 6.84 \, \text{mW} \), respectively.
The frequency of LO field is 6.947 GHz and the electric field strengths are set at the optimal working points.
The frequency detuning \( \delta f \) between the test signal and LO was +150 kHz.
We can see that, for the same applied $\mu$-wave field, all the data with the cavity are approximately 19 dB higher than those in free space. 
Compared to the Rydberg receiver in free space, the cavity-enhanced Rydberg receiver provides not only a higher sensitivity but also a larger dynamic range.
By linearly fitting the data, we can get the minimum detectable electric field at the point with $SNR=1$.
The minimum detectable electric fields are \( 168 \, \text{nV/cm} \) and \( 1.53 \, \mu\text{V/cm} \) for the cavity-enhanced and free-space receivers, respectively.
Considering the 1-Hz frequency resolution of the spectrum analyzer, the sensitivities are \( 168 \, \mathrm{nV/cm/Hz^{1/2}} \) and \( 1.53 \, \mu\mathrm{V/cm/Hz^{1/2}} \) for the two cases. 
The cavity-enhanced receiver is about 10 times more sensitive than the free-space counterpart.
The cavity-enhanced receiver can also provide a 71-dB dynamic range (from -108 dBm to -37 dBm), which is 19-dB wider than the free-space receiver (from -107 dBm to -55 dBm).
The enhancement is attributed to the optical cavity, which significantly increases the interaction between the light and the atomic system, improving the spectral contrast and thus increasing the expansion coefficient \( \kappa \).

\section{Conclusion and discussion}

We have investigated the performance of a cavity-enhanced Rydberg atomic superheterodyne receiver for precision $\mu$-wave measurements. The cavity-enhanced technique, utilizing multiple photon reflections within an optical cavity, significantly strengthens the interaction between the probe light and the atomic system, leading to notable improvements in sensitivity and signal-to-noise ratio (SNR). 
The experimental results demonstrated that, compared to the free-space case, the sensitivity and the dynamical range of the cavity-enhanced case are improved by approximately 19 dB. 

Although we have shown that the performance of the cavity-enhanced Rydberg receiver is much better than the free-space counterpart with the same conditions, the values of the absolute minimum detectable $\mu$-wave and the dynamic range are still worse than the reported free-space Rydberg receiver \cite{jing2020atomic}.
The reason is that a limited number of atoms interacts with the light field. 
In the current experiment, the beam waist of the optical cavity is 50 $\mu$m, while the Rydberg receiver invented by Jing et al. has millimeter-scale beam size. 
The interacting atom number is at least 100 times smaller in our experiment.
If a cavity with a larger mode size is adopted, the performance of the cavity-enhanced Rydberg receiver would be improved further.

\section*{Funding } This work was supported by National Key Research and Development Program of China (2021YFA1402002); National Natural Science Foundation of China (U21A6006, U21A20433, 92465201,12104277, 12104278, 12474360 and 92265108); Fund for Shanxi “1331 Project” Key Subjects; Postdoctoral Fellowship Program of CPSF (GZC20240960).
\section*{Disclosures} The authors declare no conflicts of interest.
\section*{Data availability} 
 Data underlying the results presented in this paper are not publicly available at this time but may be obtained from the authors upon reasonable request.
%%%%%%%%%%%%%%%%%%%%%%% References %%%%%%%%%%%%%%%%%%%%%%%%%

\bibliography{sample}

\end{document}